\documentstyle[12pt,amsfonts,latexsym]{article}
\begin{document}
\newcommand{\de}{\delta}\newcommand{\ga}{\gamma}
\newcommand{\e}{\epsilon} \newcommand{\ot}{\otimes}
\newcommand{\be}{\begin{equation}} \newcommand{\ee}{\end{equation}}
\newcommand{\ba}{\begin{array}} \newcommand{\ea}{\end{array}}
\newcommand{\beq}{\begin{equation}}\newcommand{\eeq}{\end{equation}}
\newcommand{\tmod}{{\cal T}}\newcommand{\amod}{{\cal A}}
\newcommand{\bemod}{{\cal B}}\newcommand{\cmod}{{\cal C}}
\newcommand{\dmod}{{\cal D}}\newcommand{\hmod}{{\cal H}}
\newcommand{\s}{\scriptstyle}\newcommand{\tr}{{\rm tr}}
\newcommand{\einsop}{{\bf 1}}
\def\oR{R^*} \def\upa{\uparrow}
\def\R{\overline{R}} \def\doa{\downarrow}
\def\dag{\dagger}
\def\ve{\epsilon}
\def\si{\sigma}
\def\ga{\gamma}
\newcommand{\reff}[1]{eq.~(\ref{#1})}
\centerline{\bf{INTEGRABILITY OF A  $t-J$ MODEL WITH IMPURITIES}}
~~~\\
\begin{center}
{\large Jon Links}
~~\\
{\em Department of Mathematics \\
University of
Queensland, Queensland, 4072,  Australia \\
e-mail jrl@maths.uq.oz.au}
~~\\
~~\\
{\large Angela Foerster}
~~\\
{\em Instituto de F\'{\i}sica da UFRGS \\ Av. Bento
Gon\c{c}alves
9500,
Porto Alegre, RS - Brazil \\ e-mail angela@if.ufrgs.br}
\end{center}
\begin{abstract}
A $t-J$ model for correlated electrons with impurities is proposed. The
impurities are introduced in such a way that integrability of the model
in one dimension is not violated. The algebraic Bethe ansatz solution of
the model is also given and it is shown that the Bethe states are
highest weight states with respect to the supersymmetry algebra
$gl(2|1)$. 
\end{abstract}
\vfil\eject
\begin{flushleft}
Corresponding author:
{\bf Dr. Jon Links} \\
Address: Department of Mathematics, \\
$\phantom{00000000}$ University of Queensland, \\
$\phantom{00000000}$ Queensland, 4072, Australia \\
e-mail: jrl@maths.uq.oz.au \\
Telephon: 61 7 336 53277 \\
Fax: 61 7 3365 1477 \\
\vspace{1cm}
\end{flushleft}
\vfil\eject
\centerline{{\bf 1. Introduction}}
~~\\

The quantum inverse scattering method (QISM) has lead to many new
results in the study of integrable and exactly solvable systems. Amongst
these is the fact that the $t-J$ model for correlated electrons is
integrable in one dimension
at the supersymmetric point $J=2t$ with the supersymmetry
algebra given by the Lie superalgebra $gl(2|1)$.  This was made
apparent in the works \cite{ek,fk} where it was shown that the
Hamiltonian could be derived from a solution of the Yang-Baxter
equation. Also,  solutions of the model
were  found by means of the algebraic Bethe ansatz.

One attractive aspect of the quantum inverse scattering method
is that one is allowed to
incorporate impurites into the system without violating integrability.
In this context, several versions of the Heisenberg chain with
impurities have been investigated \cite{joh,sse,hpe}.
For the specific case of the $t-J$ model this idea was first adopted by
Bares \cite{b} whereby the impurities were introduced into the model by
way of inhomogeneities in the transfer matrix of the system. Another
possibility was explored by Bed\"urftig et. al. \cite{bef} with
impurites given by
changing the representation of the $gl(2|1)$ generators at some lattice
sites from the fundamental three dimensional representation to the one
parameter family of typical four dimensional representations which were
introduced in \cite{bglz} to derive the supersymmetric $U$ model.

Here we wish to propose a third method for introducing integrable
impurities into the $t-J$ model. This is achieved by replacing some
lattice sites with the {\em dual} space of the fundamental three
dimensional representation. A significant point here is that only
recently have new Bethe ansatz methods been proposed in order to solve
such a system because of a lack of a suitable (unique) reference state.
Rather one is forced to work with a subspace of reference states. This
approach has been developed in the works of Abad and R{\'{\i}}os
\cite{ar1,ar2} and has already been adopted in \cite{pf} to find a Bethe
ansatz solution of the supersymmetric $U$ model starting from a
ferromagnetic space of states.

The Hamiltonian of this $t-J$ model with impurities reads
\be
H=\sum_{i=1}^Lh_{i,i+1}+\sum_{i\in I}\frac2{\lambda_i -2}h_{i,i+1}Q_i
-\frac2{\lambda_i}Q_ih_{i,i+1}  \label{ha} \ee
where
\begin{eqnarray}
h_{i,i+1}&=&-\sum_{\sigma}(c^{\dag}_{i,\si}c_{i+1,\si}+c^{\dag}_{i+1,\si}
c_{i,\si})(1-n_{i,-\si})(1-n_{i+1,-\si}) \nonumber \\
&&+2(\mathbf{S}_i.\mathbf{S}_{i+1}-\frac14n_in_{i+1})+n_i+n_{i+1}-1,
\nonumber \\
~~\nonumber \\
Q_i&=&\sum_{\si}\si(c^{\dag}_{i,\si}c_{\si}-c^{\dag}_{\si}c_{i,\si}
)(1-n_{i,-\si})(1-n_{-\si}) \nonumber \\
&&+2(\mathbf{S}_i.\mathbf{S}-\frac14n_in)-n+1
\nonumber \end{eqnarray}
and periodic boundary conditions are imposed.
Above $c_{i\pm}^{(\dagger)}$'s are spin up or down annihilation
(creation) operators, the $\mathbf{S}_i$'s
spin matrices, the $n_i$'s occupation numbers of electrons at
lattice site $i$.
The $\lambda_i$ are arbitrary complex parameters and $I$ is simply
an index set with elements in the range $1,2,...,L$. We make the
assumption that if $i\in I$ then $i\pm 1\notin I$ since otherwise extra
terms are needed in the Hamiltonian for integrability.
The operators without
site labels in the expression for $Q_i$ act on the impurity space
coupled to the site $i$. Note however that the interactions involving
the impurity sites are three site interactions involving the sites $i$
and $i+1$ as well as the impurity. The local space of states for an impurity
site has the basis
$$\left|\uparrow\right>,~~~\left|\downarrow\right>,~~~\left|\uparrow
\downarrow\right>  $$
in contrast to the local spaces for the other sites which have bases
$$\left|\uparrow\right>,~~~\left|\downarrow\right>,~~~\left|0\right>  $$
as is the case for a pure $t-J$ model. The reason for this choice
is so the Hamiltonian conserves magnetization and particle number.
Finally we mention that the first term in \reff{ha} is the Hamiltonian
for the pure $t-J$ model. We can recover this model from
\reff{ha} by taking the limit $\lambda_i\rightarrow\infty$ for each
$i\in I$.

In this paper we derive the Hamiltonian \reff{ha} by means of the
QISM which guarrantees integrabilty. We will also find solutions to the
model using the algebraic Bethe ansatz. Finally we also show that the
Bethe states which are obtained by this procedure are in fact highest
weight states with respect to the underlying supersymmetry algebra
$gl(2|1)$.

~~\\  
\centerline{{\bf 2. Derivation of the Hamiltonian} } ~\\ Recall that the Lie superalgebra $gl(m|n)$ has generators $\{E_j^i\}_{i,j=1}^{m+n}$ satisfying the commutation relations 
\be  [E^i_j,\,E^k_l]=\delta^k_jE^i_l-(-1)^{([i]+[j])([k]+[l])} 
\delta^i_lE^k_j  \label{cr}\ee
where the ${\mathbb Z}_2$-grading on the indices is determined by 
\begin{eqnarray} 
[i]&=&0 ~~\mathrm {for} ~~1\leq i\leq m,   \nonumber \\
{[i]}&=&1~~\mathrm {for} ~~ m<i\leq m+n.   \nonumber \end{eqnarray} 
This induces a ${\mathbb Z}_2$-grading on the $gl(m|n)$ generators
through
$$\left[E^i_j\right]=[i]+[j]\,(\mathrm{mod} 2). $$ 

The vector module $V$ has basis $\{v^i\}_{i=1}^{m+n}$ with action defined
by
\be E^i_jv^k=\delta^k_jv^i. \label{vr}   \ee   
Associated with this space there is a solution $R(u)\in \mathrm {End}
(V\otimes V)$ of the Yang-Baxter equation 
\be R_{12}(u-v)R_{13}(u)R_{23}(v)=R_{23}(v)R_{13}(u)R_{12}(u-v)
\label{yb} \ee
on the space $V\otimes V\otimes V$ which is given by
\be R(u)=I\otimes I -\frac 2u \sum_{i,j}e^i_j\otimes e_i^j (-1)^{[j]}. \label{rm1}
\ee
We remark that  \reff{yb} is acting on a supersymmetric space so the
multiplication of tensor products is governed by the relation
\be (a\otimes b)(c\otimes d)=(-1)^{[b][c]}ac\otimes bd  \label{mult} \ee
for homogeneous operators $b,c$. 

The solution given by \reff{rm1} allows us to construct a universal
$L$-operator which reads 
\be L(u)=I\otimes I -\frac 2u\sum_{i,j}e^i_j\otimes E^j_i (-1)^{[j]}.
\label{lop} \ee 
This operator gives us a solution of the Yang-Baxter equation of the
form
$$ R_{12}(u-v)L_{13}(u)L_{23}(v)=L_{23}(v)L_{13}(u)R_{12}(u-v)$$ 
on the space $ V\otimes V\otimes gl(m|n)$ which follows from the
commutation relations \reff{cr}.
 The dual representation to \reff{vr} acts on the module $V^*$ with
 basis $\{v_i\}_{i=1}^{m+n}$ and the action is given by 
 \be E^i_jv_k=-(-1)^{[i]+[i][j]}\delta^i_kv_j \label{dr}.  \ee
By taking this representation in the expression \reff{lop} we obtain the
following $R$-matrix 
\be \oR(u)=I\otimes I +\frac 2u\sum_{i,j} e^i_j\otimes e^i_j
(-1)^{[i][j]}  \label{rm2} \ee 
giving the solution 
\be R_{12}(u-v)\oR_{13}(u)\oR_{23}(v)=\oR_{23}(v)\oR_{13}(u)R_{12}(u-v) 
\label{yb1}  \ee  
on $V\otimes V\otimes V^*.$ 
 
We wish to construct an impurity model with generic quantum spaces
represented by $V$ and the impurity spaces by $V^*$. To this end take
some index set $I=\{p_1,p_2,....,p_l\},~1\leq p_i\leq L$ and define
$$W=\bigotimes_{i=1}^L W_i$$ 
where 
\begin{eqnarray} 
W_i&=&V~~~\,~~~~~~~\mathrm{if} ~~i\notin I,  \nonumber \\
W_i&=&V\otimes V^*~~~\mathrm{if} ~~i\in I. \label{w}
\end{eqnarray}   
In other words for each $i\in I$  we are coupling an impurity into the lattice  which will
be situated between the sites $i$ and $i+1$.

Next we define the monodromy  matrix 
$$ T(u,\,\{\lambda\})=\R_{01}(u)\R_{02}(u)....\R_{0L}(u)$$ 
where we have 
\begin{eqnarray}
\R_{0i}(u)&=&R_{0i}(u)~~~~~~~~~~~~~~~~~~~~  \mathrm{for}~~ i\notin I,   \nonumber \\
\R_{0i}(u)&=&R_{0i'}(u)\oR_{0i''}(u-\lambda_i)~~~  \mathrm{for}~~ i\in I. 
\nonumber\end{eqnarray}   
Above, the indices $i'$ and $i''$ refer to the two spaces in $W_i$ (cf.
\reff{w}) and the $\lambda_i$ are arbitrary complex parameters. 
A consequence of eqs. (\ref{yb}, \ref{yb1}) is that the monodromy
matrix satisfies the intertwining relation
\be R_{12}(u-v)T_{13}(u)T_{23}(v)=T_{23}(v)T_{13}(u)R_{12}(u-v) \label{int} 
\ee   
acting on the space $V\otimes V\otimes W$. The transfer matrix is defined by
\be \tau(u)=\mathrm{tr}_0 \sigma_0T(u) \label{tm}  \ee    
where the matrix $\sigma$ has entries
$$\sigma_j^i=(-1)^{[i][j]}\delta_j^i$$    
from which the Hamiltonian is obtained through
\be H=-2\left.\frac d{du}\ln (v^L\tau(u))\right|_{u=0}. \label{ham} \ee
In this derivation we used the property
$$ Q_i ( -2 h_{i,i+1} ) Q_i \, = Q_i $$
which follows from the fact that $Q$ projects onto a one-dimensional
space. This simplifies the calculations and is
one of the reasons why this model is much simpler
than other impurity chains.
From \reff{int} we conclude by the usual argument that the transfer
matrix provides a set of abelian symmetries for the model and hence the
Hamiltonian is integrable. In the next section we will solve the model
by the algebraic Bethe ansatz approach. The explicit form of the
Hamiltonian \reff{ha} is given by making the following
identification between the basis elements of $V$, $V^*$ and the
electronic states:
\begin{eqnarray}
v^1=\left|\uparrow\right>,&~~~~~&v_1=\left|\downarrow\right>, \nonumber
\\
v^2=\left|\downarrow\right>,&~~~~~&v_2=\left|\uparrow\right>, \nonumber
\\
v^3=\left|0\right>,&~~~~~&v_3=\left|\uparrow\downarrow\right>.
\nonumber\end{eqnarray}
~\\

\centerline{\bf {3. Algebraic Bethe ansatz solution}}
~\\

By a suitable redefinition of the matrix elements, the solutions
(\ref{rm1}, \ref{rm2})
may be written in terms of operators which satisfy the Yang-Baxter
equations (\ref{yb}, \ref{yb1})
without $\mathbb{Z}_2$-grading (see e.g. \cite{del}). These operators read
\begin{eqnarray}  R(u)&=&\sum_{i,j}e^i_i\otimes e^j_j(-1)^{[i][j]}-\frac
{2}{u} e^i_j\otimes e^j_i \nonumber\\
\oR(u)&=&\sum_{i,j}e^i_i\otimes
e^j_j(-1)^{[i][j]}+\frac{2}{u}e^i_j\otimes e^i_j
(-1)^{[i]+[j]} \nonumber \end{eqnarray}
and hereafter we will use these forms. 
In the following we will also need the $R$-matrices  
\begin{eqnarray}
r(u)&=&\sum_{i,j=2}^3(-1)^{[i][j]}e^i_i\otimes e^j_j- \frac 2u e^i_j\otimes e^j_i \nonumber \\
r^*(u)&=&\sum_{i,j=2}^3(-1)^{[i][j]}e^i_i\otimes e^j_j+ \frac 2u
e^i_j\otimes e^i_j(-1)^{[i]+[j]} \nonumber \end{eqnarray}
which belong to a $gl(1|1)$ invariant (6-vertex) system.
From this matrices  we define the monodromy matrices
\begin{eqnarray}  
t(v,\{u\})&=&r_{01}(v-u_1)r_{02}(v-u_2)...r_{0N}(v- 
u_N),    \nonumber \\ 
t^*(v,\{\lambda\})&=&r^*_{01}(v-\lambda_1)r^*_{02}(v-\lambda_2)...r^*_{0l}(v-
\lambda_l).  \nonumber \end{eqnarray}        

First we construct the Yangian algebra which has elements
$\{Y^i_j(u)\}_{i,j=1}^{m+n}$. Relations amongst these elements are
governed by the constraint
\be R_{12}(u-v)Y_{13}(u)Y_{23}(v)=Y_{23}(v)Y_{13}(u)R_{12}(u-v) \label{y}\ee
where
$$Y(u)=\sum_{i,j}e^i_j\otimes Y^j_i(u). $$ 
By comparison with \reff{int} we see that the monodromy matrix provides
a representation of this algebra acting on the module W by the mapping 
\be \pi(Y^i_j(u))^k_l=(-1)^{([i][l]+[j][l]+[i][k])}T^{ik}_{jl}(u).
\label{rep} \ee
Moreover the transfer matrix is expressible in terms of this
representation by 
$$ \tau(u)=\sum_{i=1}^3(-1)^{[i]+[i][k]}\pi(Y^i_i(u))^k_l $$ 
The phase factors present above are required since the Yangian algebra
is defined with a non-graded $R$-matrix. In the following we will omit
the symbol $\pi$ for ease of notation.

For a given $\{\alpha\}=(\alpha_1,\alpha_2,...,\alpha_l)$,
$\alpha_i=2,3$ we define the vector $v^{\{\alpha\}}\in W$ by 
$$v^{\{\alpha\}}=\bigotimes_{i=1}^L w^i  $$ 
where
\begin{eqnarray}
w^i=&v^1 ~~~~~~~~~~~&\mathrm{for}~~ i\notin I,   \nonumber \\
w^i=&v^1\otimes v_{\alpha_j} ~~~&\mathrm{for}~~ i=p_j\in I. \nonumber
\end{eqnarray}
Now set $X=\mathrm{span}~\{v^{\{\alpha\}}\}$. It is important to observe
that the space $X$ is closed under the action of the elements
$Y^i_j(u),~i,j=2,3$ which generate a sub-Yangian. We may in fact write
$$Y^i_j(u)v^{\{\alpha\}}=t^{*i\{\alpha\}}_{\,j\{\alpha'\}}(u,\{\lambda\})
v^{\{\alpha'\}}  $$
which follows from the fact that the $Y^i_j(u)~i,j=2,3$ act trivially on
the vector $v^1$ in the sense
\begin{eqnarray}
&&Y^2_2(u)v^1=Y^3_3(u)v^1=v^1 \nonumber \\
&&Y^2_3(u)v^1=Y^3_2(u)v^1=0\nonumber  \end{eqnarray}   

Setting 
$$S^{\{\beta\}}(\{u\})=Y^{\beta_1}_1(u_1)Y^{\beta_2}_1(u_2)....
Y^{\beta_N}_1(u_N),~~~~\beta_i=2,3 $$ 
we look for a set of eigenstates of the transfer matrix of the form
\be \Phi^j=\sum_{\{\beta,\alpha\}}S^{\{\beta\}}(\{u\})v^{\{\alpha\}}
F^j_{\{\beta,\alpha\}} \label{eig} \ee  
where the $F^j_{\{\beta,\alpha\}}$ are undetermined
co-efficients. We appeal to the algebraic equations given by \reff{y} to
determine the constraints on the variables $u_i$ 
needed to force \reff{eig} to be an
eigenstate.  Although many relations occur as a result of \reff{y} only
the following are required: 
\begin{eqnarray}   
Y^1_1(v)Y^{\beta}_1(u)&=&a(u-v)Y^{\beta}_1(u)Y^1_1(v)-b(u-v)T_1^{\beta}(v)Y^1_1
(u)  \label{yba1} \\   
Y^{\gamma'}_{\gamma}(v)Y^{\alpha}_1(u)&=&Y^{\alpha'}_1(u)Y^{\gamma''}_{
\gamma}(v)r^{\gamma'\alpha}_{\gamma''\alpha'}(v-u)  
-b(v-u)Y^{\gamma'}_1(v)Y^{\alpha}_{\gamma}(u)  \label{yba2}
 \\  a(v-u)Y_1^{\alpha}(v)Y_1^{\beta}(u)&=&
Y_1^{\beta'}(u)Y_1^{\alpha'}(v) r^{\beta\alpha}_{\beta'\alpha'}(v-u) 
\label{yba3} \end{eqnarray}
with $a(u)=1-2/u$ and $b(u)=-2/u$. 
All of the indices in eqs.(\ref{yba1}, \ref{yba2}) assume only the
values 2 and 3. Using \reff{yba1} two types of terms arise when $Y^1_1$
is commuted through $Y^{\alpha}_1$. In the first type $Y^1_1$ and
$Y_1^{\alpha}$ preserve their arguments and in the second type their
arguments are exchanged. The first type of terms are called {\it wanted
terms} because they will give a vector proportional to $\Phi^j$, and the
second type are {\it unwanted terms} (u.t.).    
We find that 
\be Y^1_1(v)\Phi^j=a(v)^L\prod_{i=1}^N a(u_i-v)\Phi^j + \mathrm{u.t.}. 
\label{y11} \label{can1} \ee
Similarly, for $i=2,3$ we have from \reff{yba2} (no sum on $i$)  
\begin{eqnarray}
Y^i_i(v)\Phi^j&=&S^{\{\beta'\}}(\{u\})Y^i_k(v)t^{k\{\beta\}}_{i\{\beta'\}}
(v,\{u\})v^{\{\alpha\}}F^j_{\{\beta,\alpha\}} + \mathrm{u.t.} \nonumber
\\
&=& S^{\{\beta'\}}(\{u\})
t^{k\{\beta\}}_{i\{\beta'\}}(v,\{u\})t^{*i\{\alpha\}}_{\,k\{\alpha'\}}
(v,\{\lambda\})v^{\{\alpha'\}}F^j_{\{\beta,\alpha\}}
+\mathrm{u.t} \nonumber \\
&=&
S^{\{\beta'\}}(\{u\}) 
\overline{t}^{i\{\beta,\alpha\}}_{i\{\beta',\alpha'\}}(v,\{u,\lambda\})
v^{\{\alpha'\}}F^j_{\{\beta,\alpha\}}   +\mathrm{u.t.}
\nonumber\end{eqnarray}  
where 
$$\overline{t}^{i\{\beta,\alpha\}}_{i\{\beta',\alpha'\}}(v,\{u,\lambda\})
=t^{k\{\beta\}}_{i\{\beta'\}}(v,\{u\})t^{*i\{\alpha\}}_{\,k\{\alpha'\}}
(v,\{\lambda\}). $$ 
The contribution to the eigenvalues of the transfer matrix is
\be Y^2_2(v)\Phi^j+(-1)^{1+[j]}Y^3_3(v)\Phi^j 
=\sum_{i=2}^3(-1)^{[i]+[i][j]}\overline{t}^{i\{\beta,\alpha\}}_{i\{\beta',\alpha'\}}(v,\{u,\lambda\})
S^{\{\beta'\}}(\{u\})v^{\{\alpha'\}}F^j_{\{\beta,\alpha\}}+\mathrm{u.t.}
\label{can2} \ee 
At this point we need to perform a second-level, or {\it nested} Bethe ansatz
procedure to diagonalize the matrix  
$$
\tau_1(v)^{\{\beta,\alpha\}}_{\{\beta',\alpha'\}} 
=\sum_{i=2}^3(-1)^{[i]+[i][\{\beta,\alpha\}]}\overline{t}^{i\{\beta,\alpha\}}_{i
\{\beta',\alpha'\}}(v,\{u,\lambda\}) $$   
where we have used the fact that $F^j_{\{\beta,\alpha\}}=0$ unless
$[j]=[\{\beta,\alpha\}]$.  
The above matrix is simply the transfer matrix for a $gl(1|1)$ invariant
system acting in the tensor product representation of $N$ copies of the vector
representation with inhomogeneities $\{u\}$ and $l$ copies of the dual
representation with inhomogeneities $\{\lambda\}$. 

To diagonalize this matrix we construct the Yangian generated by  
$$y(u)=\sum_{i,j=2}^3e^i_j\otimes y^j_i(u) $$ 
subject to the constraint
\be r_{12}(u-v)y_{13}(u)y_{23}(v)=y_{23}(v)y_{13}(u)r_{12}(u-v).
\label{yy} \ee 
From the above set of relations we will need the following
\begin{eqnarray}
y^2_2(v)y^3_2(u)&=&a(u-v)y^3_2(u)y^2_2(v)-b(u-v)y^3_2(v)y^2_2(u), \label{y1}\\
y^3_3(v)y^3_2(u)&=&-a(u-v)y^3_2(u)y^3_3(v)-b(v-u)y^3_2(v)y^3_3(u), 
\label{y2}\\
y^3_2(v)y^3_2(u)&=&\frac{-a(u-v)}{a(v-u)}y^3_2(u)y^3_2(v). \label{y3} \end{eqnarray}
Proceeding similarly as before, we look for eigenstates of the form
$$\phi=y^3_2(\gamma_1)y^3_2(\gamma_2)...y^3_2(\gamma_M)w $$ 
with the vector $w$ given by 
$$w=S^{\{2\}}(\{u\})v^{\{3\}}.$$  
Using (\ref{y1},\ref{y2}) it follows that 
$$\tau_1(v)\phi=\Lambda_1(v)\phi +\mathrm{u.t.} $$
with 
$$ \Lambda_1(v)=\prod_{i=1}^N a(v-u_i)\prod_{k=1}^M
a(\gamma_k-v)-\prod_{j=1}^la(v-\lambda_j)\prod_{k=1}^Ma(\gamma_k-v)$$ 
The unwanted terms cancel provided the parameters $\gamma_k$  satisfy 
the Bethe ansatz equations (BAE)
\be \prod_{i=1}^Na(\gamma_k-u_i)=\prod_{j=1}^la(\gamma_k-\lambda_j), 
~~~~k=1,2,...,M. \label{bae1}\ee
Combining this result with \reff{y11} we obtain for the eigenvalues of
the transfer matrix  \reff{tm}
\be \Lambda(v)=a(v)^L\prod_{i=1}^Na(u_i-v)+\Lambda_1(v). \label{eigv}\ee 
Cancellation of the unwanted terms in (\ref{can1},\ref{can2}) leads to a second set of BAE
which are   
\be a(u_h)^L\prod_{i=1}^N\frac{a(u_i-u_h)}{a(u_h-u_i)}=-
\prod_{k=1}^Ma(\gamma_k-u_h) ~~~~h=1,2,....,N.  \label{bae2} \ee 
We will not give the details proving the cancellation of the unwanted
terms but remark that the calculation is analogous to that given in
\cite{fk} for the pure $t-J$ chain. 

Making a change of variable $u\rightarrow iu+1,~\gamma\rightarrow
i\gamma +2,~\lambda\rightarrow i\lambda +1$ the BAE read
\begin{eqnarray}
-\left(\frac{u_h+i}{u_h-i}\right)^L&=&\prod_{i=1}^N\frac{u_h-u_i+2i}{
u_h-u_i-2i}
\prod_{k=1}^M\frac{u_h-\gamma_k-i}{u_h-\gamma_k+i},~~h=1,....,N,~~~
\label{be1} \\
\prod_{i=1}^N\frac{\gamma_k-u_i+i}{\gamma_k-u_i-i}
&=&\prod_{j=1}^l\frac{\gamma_k-\lambda_j+i}{\gamma_k-\lambda_j-i},
~~~~~~~~~~~~~~~~~~~~~~~~k=1,....,M .~~~  \label{be2} \end{eqnarray}
In the absence of impurities (limit $l \rightarrow 0$) we recover
the form of the BAE first derived by Sutherland \cite{su}
and later by Sarkar \cite{sar} for the usual t-J model.
Adopting the string conjecture, or more specifically assuming that the
solutions $u_i$ are real or appear as complex conjugate pairs and the
$\lambda_j$ are real, we find
string solutions
$$u_{\alpha\beta}^n=u^n_{\alpha}+i(n+1-2\beta),~~~\alpha=1,2,...,N_n,~~
\beta=1,2,...,n,~~n=1,2,...  $$   
and the $\gamma_k$ are real. 
The number of $n$-strings $N_n$ satisfy the relation
$$N=\sum_nnN_n.$$    

As was shown in the papers \cite{ek,fk} two other forms of the
Bethe ansatz  exist which are obtained by choosing a different
grading for the indices of the $gl(2|1)$ generators.
Recall that the above calculations were performed with the choice
$$[1]=[2]=0,~~~~[3]=1. $$ 
Choosing 
$$[1]=1,~~~~[2]=[3]=0 $$ 
yields the eigenvalue expression
\begin{eqnarray}
\Lambda(v)&=&-a(-v)^L\prod_{i=1}^Na(v-u_i)+\prod_{k=1}^M
a(v-\gamma_k) \prod_{j=1}^la(\lambda_j-v)  \nonumber \\
&&+\prod_{i=1}^Na(v-u_i)\prod_{k=1}^M a(\gamma_k -v)\nonumber  \end{eqnarray}
subject to the BAE
\begin{eqnarray}
a(-u_i)^L&=& \prod_{k=1}^Ma(\gamma_k-u_i),
~~~~i=1,2,...,N  \nonumber \\
\prod_{k=1}^M\frac{a(\gamma_h-\gamma_k)}{a(\gamma_k-\gamma_h)} 
&=&-\prod_{i=1}^Na(\gamma_h-u_i)\prod_{j=1}^l\frac1{a(\lambda_j-\gamma_h)}
,~~~~h=1,2,...,M.\nonumber 
\end{eqnarray}
In the limit $l \rightarrow 0$ we recover Lai's \cite{lai}  (see also
\cite{schlot} )
Alternatively, choosing
$$[1]=[3]=0,~~~~[2]=1 $$
yields the eigenvalue expression
\begin{eqnarray}
\Lambda(v)&=&a(v)^L\prod_{i=1}^Na(u_i-v)+\prod_{k=1}^Ma(v-\gamma_k)
\prod_{j=1}^la(\lambda_j-v)   \nonumber \\
&&-\prod_{i=1}^Na(u_i-v)\prod_{k=1}^M
a(v-\gamma_k) \nonumber  \end{eqnarray} 
with the BAE
\begin{eqnarray}
a(u_i)^L&=& \prod_{k=1}^Ma (u_i-\gamma_k)
~~~~i=1,2,...,N, \nonumber \\
\prod_{i=1}^Na(u_i-\gamma_k)&=&\prod_{j=1}^la(\lambda_j-\gamma_k) ~~~~
k=1,2,...,M. \nonumber \end{eqnarray} 

Finally, from the definition of the Hamiltonian \reff{ham} we see that
the energies are given by 
$$E=-2\left.\frac{d}{dv} \ln\left(v^L\Lambda(v)\right)\right|_{v=0}. $$ 
Using the eigenvalue expression \reff{eigv} we obtain 
$$E=L+4\sum_{i=1}^N\frac{1}{1+u_i^2}$$ 
where the $u_i$ are solutions to the equations (\ref{be1},\ref{be2}). 
~~\\~~\\
\centerline{\bf{4. Highest weight property}}
~~\\
Next we wish to show that the eigenstates constructed in the previous
section are in fact highest weight states with respect to the underlying
supersymmetry algebra $gl(2|1)$. The highest weight property of the
Bethe states has been proved for many models, such as the 
Heisenberg chain \cite{fad} and its generalized version \cite{kir2}, 
the Kondo model \cite{kir1}, the usual $t-J$ model \cite{fk}, 
the Hubbard chain \cite{fab1,fab2,fab3} and its $gl(2/2)$ extension \cite{schout}.
However, as far as we are aware it has never been shown
before in the case where a subspace of reference states has been used
in the Bethe ansatz procedure.

Let us begin by considering 
$$E^2_3\Phi^j=\sum_{\{\beta,\alpha\}}E^2_3S^{\{\beta\}}(\{u\})v^{\{\alpha
\}}F^j_{\{\beta,\alpha\}}. $$
By means of the nesting procedure we know that the co-efficients $F^j_{
\{\beta,\alpha\}}$ are such that we have the following
identification of states 
$$S^{\{\beta\}}(\{u\})v^{\{\alpha\}}F^j_{\{\beta,\alpha\}} 
=y^3_2(\gamma_1)y^3_2(\gamma_2)....y^3_2(\gamma_M)w $$ 
for a suitable solution of the BAE. By comparing eqs.
(\ref{lop},\ref{yy},\ref{rep}) it is possible to determine algebraic
relations between the elements of the Yangian algebra and the
supersymmetry algebra. For our purposes we need the following 
\be [E^2_3,\,y^3_2(u)]^{\alpha}_{\beta}=-y^2_2(u)^{\alpha}_{\beta}+
y^3_3(u)^{\alpha}_{\beta}(-1)^{[\alpha]}. \label{tran1} \ee   
Noting that $E^2_3w=0$ it is evident that we may write
$$E^2_3y^3_2(\gamma_1)....y^3_2(\gamma_M)w=\sum_{h=1}^Mx_hX_h $$
with 
$$X_h=y^3_2(\gamma_1).....y^3_2(\gamma_{h-1})y^3_2(\gamma_{h+1})....y^3_2
(\gamma_M)w $$ 
and the $x_h$ some yet to be determined co-efficients. 
To find $x_h$ we write
$$y^3_2(\gamma_1)....y^3_2(\gamma_M)w=\prod_{j=1}^{h-1}\frac{-a(\gamma_h-
\gamma_j)}{a(\gamma_j-\gamma_h)} y^3_2(\gamma_h)X_h$$
where we have used \reff{y3}. Now by using the relations 
(\ref{y1},\ref{y2},\ref{tran1}) and looking only for those terms which
give a vector proportional to $X_i$ we find that
$$x_h=\prod_{j=1}^{h-1}\frac{-a(\gamma_h-\gamma_j)}{a(\gamma_j-\gamma_h)}
\left(\prod_{j=1}^la(\gamma_h-\lambda_j)\prod_{k\neq h}^Ma(\gamma_k-\gamma_h)
-\prod_{i=1}^Na(\gamma_h-u_i)\prod_{k\neq h}^Ma(\gamma_k-\gamma_h) \right) $$
which vanishes because of \reff{bae1}. Thus we see that 
$$E^2_3\Phi^j=0.$$

Next we consider the action of $E^1_2$ on $\Phi^j$. 
Using eqs. (\ref{lop},\ref{y},\ref{rep}) we find the
commutation relation 
\be [E^1_2,\,Y^{\alpha}_1(u)]=\delta^{\alpha}_2Y^1_1(u)-Y^{\alpha}_2(u).
\label{tran2} \ee   
As before, since $E^1_2v^{\{\alpha\}}=0$ we can write the general
expression
$$ E^1_2\Phi^j=\sum_{h,\beta} z_{h,\beta}Z_{h,\beta} $$ 
where 
$$Z_{h,\beta}=S^{\{\beta_h^-\}}(\{u_h^-)\})S^{\{\beta_h^+\}}(\{u_h^+)\}
)v^{\{\alpha\}}F^j_{\{\beta,\alpha\}}    $$
and for any vector $\{w\}$ we have 
$$\{w_h^-\}=(w_1,w_2,...,w_{h-1}),~~\{w_h^+\}=( w_{h+1},.....,w_N).
$$  
To calculate $z_{h,\beta}$ we begin by writing
\begin{eqnarray}
\Phi^j&=&S^{\{\beta_h^-\}}(\{u_h^-\})Y^{\beta_h}_1(u_h)S^{\{\beta_h^+\}}(\{u_h^+\})v^{\{\alpha\}}F^j_{\{\beta,\alpha\}}  \nonumber \\
&=&\prod_{i=1}^{h-1}a(u_i-u_h)^{-1}t^{\beta_h\{\beta_h^-\}}_{
\gamma\{\gamma_h^-\} }(-u_h,\{-u_h^-\})Y^{
\gamma}_1(u_h)S^{\{\gamma_h^-\}}(\{u_h^-\})S^{\{\beta_h^+\}}(\{u_h^+\}
)v^{\{\alpha\}}F^j_{\{\beta,\alpha\}}  
\nonumber \end{eqnarray} 
where we have used the relation \reff{yba3}. 
Now applying \reff{tran2} and using the relations (\ref{yba1},\ref{yba2}) 
to determine the terms which
give a vector proportional to $Z_{h,\beta}$ we find that 
$$z_{h,\beta}=\delta_2^{\beta_h}\left(a(u_h)^L\prod_{i\neq
h}^Na(u_i-u_h) -\prod_{i\neq
h}^Na(u_h-u_i)\prod_{k=1}^Ma(\gamma_k-u_h)\right) $$ 
which vanishes as a result of \reff{bae2}. We then conclude that
$$E^1_2\Phi^j=0$$ 
which completes the proof that the Bethe states are $gl(2|1)$ highest
weight states. We observe that this property can also be proved
for the other two choices of gradings in a similar way.
~~\\
~~\\
\centerline{{\bf 5. Conclusions}}
~~\\

In this paper we have introduced a new integrable version of the
t-J model with impurities. The model was solved through an algebraic
Bethe ansatz method and three different forms of the BAE were derived. A
proof of the highest weight property of the Bethe vectors with respect
to the $gl(2|1)$ superalgebra was also presented. A possible application
of these results would be an analysis of the structure of the ground
state and some low lying excitations of the model in the thermodynamic
limit.

~~\\
~~\\
\centerline{{\bf Acknowledgements}}
~~\\

JL is supported by an Australian Research Council Postdoctoral Fellowship.
AF thanks CNPq-Conselho Nacional de Desenvolvimento Cient\'{\i}fico e
Tecnol\'ogico for financial support.


\end{document}